\documentclass[sigconf, nonacm]{aamas} 

\usepackage{balance} 
\usepackage{times}
\usepackage{helvet}
\usepackage{courier}
\usepackage{microtype}
\usepackage{caption}
\usepackage{subcaption}
\usepackage{booktabs} 
\usepackage{multirow}
\usepackage{amsmath}

\usepackage{amssymb}
\usepackage{algorithm}
\usepackage[noend]{algpseudocode}
\usepackage{xcolor}
\usepackage[utf8]{inputenc}
\DeclareMathOperator*{\argmax}{arg\,max}

\title{Learning Emergence of Interaction Patterns across Independent RL Agents in Multi-Agent Environments}

\author{Vasanth Reddy Baddam}
\affiliation{
  \institution{Virginia tech}
  \city{Arlington, VA}
  \country{United States}}
\email{vasanth2608@vt.edu}

\author{Suat Gumussoy}
\affiliation{
  \institution{Siemens Technology}
  \city{Princeton, NJ}
  \country{United States}}
\email{suat.gumussoy@siemens.com }

\author{Almuatazbellah Boker}
\affiliation{
  \institution{Virginia tech}
  \city{Arlington, VA}
  \country{United States}}
\email{boker@vt.edu}

\author{Hoda Eldardiry}
\affiliation{
  \institution{Virginia tech}
  \city{Blacksburg, VA}
  \country{United States}}
\email{hdardiry@vt.edu}

\begin{abstract}
Many real-world problems, such as controlling swarms of drones and urban traffic, naturally lend themselves to modeling as multi-agent reinforcement learning (RL) problems. However, existing multi-agent RL methods often suffer from scalability challenges, primarily due to the introduction of communication among agents. Consequently, a key challenge lies in adapting the success of deep learning in single-agent RL to the multi-agent setting. In response to this challenge, we propose an approach that fundamentally reimagines multi-agent environments. Unlike conventional methods that model each agent individually with separate networks, our approach, the Bottom Up Network (BUN), adopts a unique perspective. BUN treats the collective of multi-agents as a unified entity while employing a specialized weight initialization strategy that promotes independent learning. Furthermore, we dynamically establish connections among agents using gradient information, enabling coordination when necessary while maintaining these connections as limited and sparse to effectively manage the computational budget. Our extensive empirical evaluations across a variety of cooperative multi-agent scenarios, including tasks such as cooperative navigation and traffic control, consistently demonstrate BUN's superiority over baseline methods with substantially reduced computational costs. 
\end{abstract}


\begin{document}


\pagestyle{fancy}
\fancyhead{}


\maketitle 

\section{Introduction}
Multi-Agent Reinforcement Learning (MARL) is an exciting field within artificial intelligence and machine learning. It has become increasingly crucial for tackling real-world problems where multiple autonomous agents interact in complex environments. However, MARL can be computationally expensive due to agents needing to interact with each other. On the other hand, Independent learning algorithms have practical advantages. They require fewer computational resources and can scale to larger environments. However, they face challenges in multi-agent settings due to non-stationarity, which can affect traditional reinforcement learning guarantees. \par 
A recent study by~\cite{lee2022investigation} demonstrated that independent learning algorithms can perform exceptionally well in cooperative scenarios when individual agents have full observability. However, full observability is often computationally expensive. In many real-world applications with multiple agents, interactions among agents are infrequent. For example, think of two robots delivering documents in an office. They do not need to closely monitor each other often, only when they have to pass through the same door. Similarly, a group of drones might be sent on a rescue mission where communication is challenging. They cannot always share information, so they need to be innovative and cooperate only when necessary. \par 
We ask a fundamental question: "When is coordination essential?" and if needed, "How infrequent can interactions be?". Our approach aims to use local interactions, allowing agents to act independently as much as possible and keeping communication minimal. This paper focuses on cooperative MARL scenarios in partially observable environments where agents have varying observation abilities. \par 
This paper considers MARL scenarios wherein the task is cooperative and agents are situated in a partially observable environment. However, each is endowed with different observation power. This paper aims to tackle this problem by reimagining the multi-agent setting by redefining the representation of the multi-agent system, treating it as a single agent with a unique neural network weight initialization scheme. This scheme ensures decentralized operation, allowing each agent to act locally and independently without interacting with other agents during execution. Subsequently, we leverage gradient information among agents to establish connections, enabling coordination when necessary. Importantly, we maintain these connections as limited and sparse. \par 
\textbf{Contributions : }In this paper, we introduce an approach inspired by the bottom-up approach described from our previous work~\cite{reddy2024searching}, Bottom Up Network, denoted as BUN, distinguished by a unique weight initialization strategy that is sparse and decentralized. Unlike conventional dense-weight initialization methods, BUN adopts a sparse initialization approach, ensuring that agents make independent local decisions while minimizing computational costs due to its sparsity. This initialization fosters local decision-making among agents, and weights emerge using gradient information during training. This weight emergence not only fosters coordination but also reveals the topology of the agents in the environment. Through extensive empirical evaluations in cooperative multi-agent scenarios, our approach showcases the advantages of sparse and decentralized initialization, significantly contributing to multi-agent reinforcement learning by addressing non-stationarity challenges and promoting efficient decentralized decision-making and communication among agents.\par 
In our evaluation of BUN, we test it in two applications: Cooperative Navigation and Traffic Signal Control environments. Our experiments consistently show that BUN outperforms baseline methods. Notably, it achieves performance levels similar to methods with full communication while using significantly fewer computational resources. It is important to note that BUN complements existing cooperative multi-agent algorithms, making them more practical and cost-effective for real-world applications.

\section{Related work} \label{relatedwork}
Multi-agent Reinforcement Learning (MARL) has garnered significant attention, with Independent Q-learning (IQL) being one of the pioneering approaches. IQL trains distinct Q-value functions for each agent, assuming that other agents remain static components within the environment. However, this method faces a notable challenge when the environment becomes non-stationary, leading to instability issues, mainly when applied to large-scale systems. Recent research~\cite{lee2022investigation} has illuminated scenarios in which IQL can demonstrate effectiveness within multi-agent systems by relying solely on local observations, eliminating the necessity for coordination. Similarly, IA2C and IPPO have shown promise in specific environments where agent coordination may not be necessary despite the non-stationarity concern. It is worth noting that in cases where learned behaviours cannot replace coordination, direct communication between agents remains indispensable. \par 
Recent works have focused on training actor-critic algorithms to address the challenges posed by non-stationarity and coordination. In these approaches, the critic is centralized and utilizes global information during training, while actors employ local information during execution. MADDPG~\cite{lowe2017multi} learns a centralized critic for each agent by providing all agents' joint state and actions to the critic. Policies for each agent are trained using the DDPG algorithm~\cite{lillicrap2015continuous}. COMA~\cite{foerster2018counterfactual} also employs a centralized critic but estimates a counterfactual advantage function to assist with multi-agent credit assignment, isolating the impact of each agent's actions. VDN~\cite{sunehag2017value} decomposes a centralized state-action value function into a sum of individual agent-specific functions. However, this decomposition imposes a strict prior, which needs to be thoroughly justified and limits the complexity of the agents' learned value functions.
Q-Mix~\cite{rashid2020monotonic} builds upon this by removing the need for additive decomposition of the central critic, instead imposing a less restrictive monotonicity requirement on agents' state-action value functions. This approach allows for a learnable mixing of individual functions without restricting the complexity of the functions that can be learned. Nevertheless, these methods do not leverage the structural information inherent in the environment. In these approaches, each agent's observation typically consists of a concatenation of the states of other agents and various environmental features. Additionally, using a centralized critic prevents the learned policy from generalizing to scenarios with fewer agents than encountered during training. Moreover, granting access to the global state for a dense centralized critic incurs significant computational costs. To tackle these challenges, several prior studies have explored the learning of communication protocols that exploit the underlying environmental structure~\cite{foerster2016learning, sukhbaatar2016learning, peng2017multiagent}. These studies train multiple agents to acquire a communication protocol and have demonstrated that communicating agents achieve enhanced rewards across various tasks. Other approaches, such as~\cite{das2019tarmac, jiang2019graph, agarwal2019learning}, utilize a dot-product attention mechanism for inter-agent communication, restricting communication to an agent's neighbours to mitigate computational overhead. However, learning communication protocols across agents using such methods can be computationally intensive. In our work, we aim to address this challenge by focusing on learning shared parameters that are as sparse as possible.\par
Recently, a notable interest has been in training sparse neural networks within Deep Reinforcement Learning (DRL). In~\cite{livne2020pops}, they proposed the Policy Pruning and Shrinking (PoPs) method. This approach incorporates iterative policy pruning as an intermediary step to guide the dimensions of dense neural networks specifically tailored for DRL agents. Turning to exploring the Rigged Lottery Ticket (RigL) Hypothesis in DRL, the work in~\cite{evci2020rigging} investigated this phenomenon. In contrast, the work in~\cite{vischer2021lottery} made substantial strides by demonstrating the discovery of sparse "winning tickets" through behaviour cloning (BC), offering invaluable insights into the initiation of sparse networks. In the research~\cite{evci2020rigging}, a spotlight was cast on the challenges associated with training DRL agents using sparse neural networks from the inherent training instability within such networks. Building upon this foundation, the work in~\cite{sokar2021dynamic} delved deeper into the issue of training instability in sparse DRL agents. Their work not only shed light on the challenges but also underscored the inherent limitations of achieving stable training in these networks. Their work response introduced dynamic sparse training, the SET algorithm, which enables end-to-end training of sparse networks within actor-critic algorithms, ultimately achieving a reasonable sparsity level.
Further advances in the field were made in~\cite{lee2018snip, arnob2021single}. These research works focused on training sparse neural networks from the ground up, eliminating the reliance on pre-trained dense models. In this work~\cite{lee2018snip}, they proposed an approach involving block-circulant masks in the early stages of training, significantly enhancing pruning efficiency in TD3 agents. While in~\cite{arnob2021single},  they introduced a unique paradigm by applying one-shot pruning algorithms in offline reinforcement learning (RL) settings. Conducting an exhaustive investigation into the various methods employed in this domain, the work~\cite{graesser2022state} conducted a comprehensive comparative study. Their work emphasized the effectiveness of pruning and highlighted the substantial improvements attained compared to conventional static sparse training techniques. However, it is worth noting that most of these works primarily revolve around DRL algorithms tailored to single-agent environments. In our paper, we venture beyond these boundaries by applying these methodologies to tackle the challenges of sparsity in multi-agent environments. Additionally, we address the dynamic determination of network topology during sparse training, adding a novel dimension to this evolving field of research.

\section{Background}
\subsection{Reinforcement Learning}
\subsubsection{Single Agent}
RL is a subfield of machine learning that focuses on training agents to make optimal decisions in sequential and continuous environments. In RL, an agent generally interacts with an environment over a series of discrete time steps. The agent perceives the current state of the environment and takes actions to influence the environment. The environment, in turn, responds to the agent's actions by transitioning to a new state and providing a reward signal. The state space represents all possible situations or configurations of the environment. The action space represents all possible actions that the agent can take. The agent's task is to learn a policy, which is a mapping from states to actions, that maximizes the expected cumulative reward over time. This can be formalized as a Markov Decision Process(MDP). MDP is defined as tuple, $\langle\mathcal{S}, \mathcal{A}, P, r, \gamma\rangle$, where $\mathcal{S}$ is the state space, $\mathcal{A}$ is the action space, $P: \mathcal{S} \times \mathcal{A} \rightarrow \Delta(S)$ defines the transition dynamics $, r: \mathcal{S}  \times \mathcal{A} \rightarrow \mathbb{R}$ is the reward function, and $\gamma \in[0,1)$ is a discount factor.

The policy $\pi: \mathcal{S} \rightarrow \Delta(\mathcal{A})$ is the strategy that the agent employs to select actions $a \in \mathcal{A} $based on the current state $s \in \mathcal{S}$. It defines the agent's behaviour and can be deterministic or stochastic. The environment rewards the agent $r$ signal after each action. The agent's objective is to learn a policy that leads to actions that maximize the total cumulative reward over time. Sometimes, costs can be used instead of rewards, and the goal becomes minimizing the cumulative cost. 
A policy $\pi$ formalizes an agent's behaviour and the associated value function $V^\pi: \mathcal{S} \rightarrow \mathbb{R}$ defined as:
$
V^\pi(s):=\mathbb{E}_{a \sim \pi(x)}\left[\mathcal{R}(x, a)+\gamma \mathbb{E}_{x^{\prime} \sim \mathcal{P}(x, a)} V^\pi\left(x^{\prime}\right)\right]
$ and state-action value functions $Q^\pi$ : $\mathcal{S} \times \mathcal{A} \rightarrow \mathbb{R}$ as: $Q^\pi(s, a):=\mathcal{R}(s, a)+\gamma \mathbb{E}_{x^{\prime} \sim \mathcal{P}(s, a)} V^\pi\left(x^{\prime}\right)$.

\subsubsection{Multi-Agent Reinforcement Learning}
Multi-agent reinforcement Learning (MARL) is an extension of single-agent reinforcement learning that deals with scenarios where multiple autonomous agents interact within a shared environment. Each agent seeks to learn a policy that maximizes its expected cumulative reward over time while considering the actions and strategies of other agents. This field is critical when addressing problems involving coordination, competition, and collaboration among multiple agents. We formalize MARL using DEC-POMDP~\cite{oliehoek2016concise}, a generalization of MDP to allow distributed control by multiple agents who may be incapable of observing the global state. A DEC-POMDP is described by a tuple $\langle\mathcal{S}, \mathcal{A}, \mathcal{R}, P, s, \mathcal{O}, \gamma\rangle$. The joint state space $\mathcal{S}$ encapsulates the collective configuration of all agents and the environment. Similarly, a joint action space $\mathcal{A}$ includes all possible combinations of actions that each agent can take. The interaction between agents and the environment unfolds over discrete time steps. Each agent $i \in \mathcal{N}$ chooses an action $a_i \in \mathcal{A}$, forming a joint action vector $\boldsymbol{a}=\left[a_i\right] \in \mathcal{A}^n$ and has partial observations $o_i \in s$. The agents' joint observations $s = (\mathbf{o}_1, \ldots, \mathbf{o}_N)$ provide insights into the collective state $\mathcal{S}$, but full knowledge of the environment is often obscured. Similarly, the agent's reward is given by the $r(o_i, a_i) \in \mathcal{R}$. Each agent $i$ takes action $a_i$ based on its own policy $\pi^i\left(a_i \mid o_i\right)$. The agents' policies $\pi_i$ now map observations to actions. The joint value function $V^\pi$ captures the expected cumulative reward achievable under the joint policy $\pi$. It is defined similarly to before, considering observations instead of states:
$
V^\pi(\mathbf{s}) = \mathbb{E}_{\mathbf{a} \sim \pi(\mathbf{a} | \mathbf{s})} \left[ \sum_{t=0}^\infty \gamma^t \sum_{i=1}^N r_i(\mathbf{s}_t, \mathbf{a}_t) \Big| \mathbf{s}_0 = \mathbf{o} \right]$. The joint state-action value function $Q^\pi$ also adapts to partial observability:
$
Q^\pi(\mathbf{s}, \mathbf{a}) = \sum_{t=0}^\infty \gamma^t \mathbb{E}_{\mathbf{s}', \mathbf{a}' \sim P(\mathbf{s}', \mathbf{a}' | \mathbf{s}, \mathbf{a})} \left[ \sum_{i=1}^N r_i(\mathbf{s}_t, \mathbf{a}_t) \right]
$. In summary, multi-agent reinforcement learning with partial observations introduces the complexity of limited information, requiring agents to adapt their policies to make effective decisions based on their local observations. Adapting value functions and policies to the observation space enables agents to handle partial observability and learn optimal strategies in challenging environments.
\subsubsection{DQN}
Deep Q-Networks (DQN)~\cite{mnih2013playing} are a class of reinforcement learning algorithms that combine Q-learning with deep neural networks, which are highly effective for function approximation. At its core, DQN aims to approximate the optimal action-value function, denoted as $Q^\pi_\theta(\mathbf{s}, \mathbf{a})$. The weight parameters $\theta$ are trained using a temporal difference loss from transitions sampled from experience replay buffer $\mathcal{D}$:
\begin{equation} \label{objective function}
    \mathcal{L}(\theta)=\mathbb{E}_{\left(s, a, r, s^{\prime}\right) \sim \mathcal{D}}\left[Q_\theta(s, a)-\left(r+\gamma \max _{a^{\prime} \in \mathcal{A}} Q_{\bar{\theta}}\left(s^{\prime}, a^{\prime}\right)\right)\right]
\end{equation}
where $\bar{\theta}$ are the weight parameters of the target network. The target network is a copy of the online network and is used to estimate the Q-values in the target. It helps in stabilizing the training process by providing a consistent target for Q-value approximation. 

\section{BUN : Bottom Up Network} \label{method}
\begin{figure}
    \begin{center}
    \includegraphics[scale=0.3]{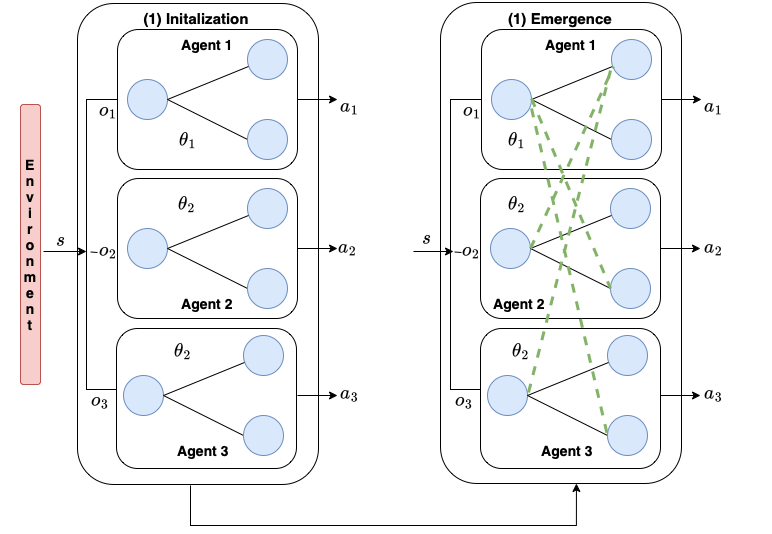}
    \end{center}
    \vspace{-3mm}
    \caption{The BUN approach involves a two-step. 1. Weight Initialization: Weights are initialized so that $i^{th}$ agent’s observation $o_i$ is directly mapped to its action $a_i$ without any dependence on the other agent’s observation. 2. Weight Emergence: We then grow the weights across the agents according to the highest magnitude gradient signal. The Green dotted line represents the newly emerged weights/connections.}
    \label{bottomuparchitecture}
\end{figure}

In practical cooperative scenarios, agents are often distributed, each equipped with its local observations, actions, and local objective rewards. Although separated within their individual networks, these agents can communicate over a shared medium to work collaboratively towards achieving global objectives. In this paper, we introduce a multi-agent reinforcement learning problem formulation, treating it as a single-agent problem with multi-discrete actions as illustrated in Figure~\ref{bottomuparchitecture}. Our methodology leverages a straightforward yet effective DQN (Deep Q-Network) algorithm. However, we can use any standard reinforcement learning algorithm with the provided network initialization. BUN starts with a sparse network, and at regularly spaced intervals, new connections emerge using gradient information. After updating the connectivity, training continues with the updated network until the next update. The main parts of our algorithm, Network Initialization, Main Objective, Weight Emergence, Update Schedule, and the various options considered for each, are explained below.  \par
\textbf{(1) Neural Network Initialization $(\theta^0)$}  As shown in Fig~\ref{bottomuparchitecture}, 
We have a single network. The input $s$ for the network is the list of all agent's observations, $s = \left[o_1, o_2, \cdots o_\mathcal{N}\right]$. The given set of states $s$ output is the list of individual actions, $a = \left[a_1, a_2, \cdots a_\mathcal{N}\right]$. 
The core logic behind BUN is based on the following principles: \textbf{1. Block Diagonal Weight Initialization $(\theta_i)$:} In BUN, we initialize the neural network's weights following a block diagonal pattern. This approach allocates specific network components to handle particular agent interactions or tasks. By incorporating agent-specific information directly into the architecture, each agent benefits from dedicated network components, fostering specialization and targeted learning.

\textbf{2. Zero Initialization for Off-Diagonal Weights $(\theta_{ij} = \mathbf{0})$:}  In contrast to the block diagonal weights, BUN initializes off-diagonal weights to zero. This design choice promotes isolation and independence between agents when they do not have direct interactions. By setting these weights to zero, we encourage agents to act autonomously when their actions do not significantly impact or depend on each other, facilitating efficient localized decision-making. \par 
\textbf{(2) Objective}
In the case of collaborative multi-agent reinforcement learning, the multi-agents have a shared reward so that there is a global objective function that must be optimized based on the collaborative efforts of the agents as given in~\eqref{objective function}. However, in our work, we need to optimize the objective function while adhering to a predefined budget of network weights. The optimization problem can be mathematically formulated as follows:
\begin{equation} \label{mainobjectivestructured}
\begin{array}{cc} 
    \text{minimize} &  \quad \mathcal{L}(\theta)\\
    \text{subject to} & \|\theta\|_{\ell_0} = b + \|\theta^0\|_{\ell_0} 
\end{array}
\end{equation} 
where $b$ is the budget, represents the limit on the number of additional network parameters that can be added to $\theta^0$. The incorporation of the budget constraint $\|\theta\|_{\ell_0} = b + \|\theta^0\|_{\ell_0}$ into the objective function encourages the emergence of weights in the network while staying within the specified budget. The parameter $b$ allows control over the degree of the sparsity: a larger $b$ tends to make the network denser, while $b=0$ corresponds to the initial sparse network initialization. Overall, this optimization problem aims to strike a balance between minimizing the loss function and controlling the sparsity of the controller $F$ within the defined budget constraints. \par 
\textbf{(3) Weight Emergence}
Solving the optimization problem~\eqref{mainobjectivestructured} directly is hindered by the presence of the sparsity constraint. This constraint introduces non-convexity and combinatorial complexity, primarily because the $\ell_0$ norm is used to quantify sparsity. To overcome these challenges, heuristic methods offer an effective approach, and in this context, a Greedy Coordinate Descent algorithm~\cite{dhillon2011nearest} proves particularly advantageous. The key idea is to iteratively select network parameters that minimize the optimization objective while adhering to the sparsity constraint. The same is explored in~\cite{evci2020rigging, yoon2017lifelong} to grow the connections across the neurons. We follow the same greedy principle to allow the emergence of the weights. During training, weights emerge across the agents based on the highest magnitude gradient rule, represented as ${(i, j)} = \argmax(|\nabla_{\theta_{ij}} \mathcal{L}(\theta)|, k)$. Here, $\theta_{ij}$ refers to the weights across the agent blocks. Given the initial zero initialization of weights, the newly emerged weights initially do not influence the network's output. Selecting weights with the highest gradients ensures a substantial reduction in the loss function during training. \par 
\textbf{(3) Update Schedule}
The connections emerge based on a predefined schedule determined by the following parameters. $b: $ The number of additional connections need to emerge. $\Delta T:$ The frequency at which the connections need to emerge, $k:$ The number of connections that need to emerge at each update, $T_{start}$ The iteration at which connection emergence should start. We allow the uniform emergence of $k$ connections at each update, sampled at a $\Delta T$ frequency until we reach the budget $b$. This gradual emergence fosters stability and reliability by preventing abrupt and potentially unstable changes in agent interactions. In addition, it makes it easier to interpret the learning process and can gain insights into how the agents adapt to new connections and make sense of the learning dynamics. The pseudo algorithm for BUN is given in Algorithm~\ref{bunalgorithm}. \par 
\textbf{Dimensionality }When using a single network, both observation and action space can grow exponentially. For example, if we use a DQN as the base network, the number of output nodes will typically grow exponentially, depending on the number of agents. One way to deal with the exponential growth in the joint action space is to use DDPG with Gumbel-Softmax action selection if the environment is discrete to avoid the exploding number of input nodes of the observation space, as well as an exploding number of output nodes of the action space. Under this paradigm, the input and output nodes only grow linearly with the number of agents, as the output nodes of a neural network in DDPG are the chosen joint action, as opposed to a DQN, where the output nodes must enumerate all possible joint actions. 

\begin{algorithm}
\caption{BUN}\label{bunalgorithm}
\begin{algorithmic}[1]
\State Initalize Network : training network $Q_{\theta}$, target network $Q_{\bar{\theta}}$, budget: $b$, Schedule : $k$, $T_{start}$, $\Delta T$ , $T_{end}$ 
\State Neural Network Initialization $\theta$
\For{$t$ in $T_{total}$}
\State Sample a batch $<s, a, r, s^{\prime}>$ from $\mathcal{D}$
\If {$t\%$ $\Delta T == 0$ $\text{and}$ $ T_{start} < t < T_{end}$
}
\For{each layer $l$}
\State ${i, j} = \argmax(|\nabla_{\theta_{ij}^{l}} \mathcal{L}(\theta)|, k)$
\State Weight Emergence through update; $\theta_{ij}^{l}$
\EndFor
\Else
\State $\theta = \theta - \alpha \nabla_{\theta_{ij}}$
\EndIf
\State Update the target network; $\bar{\theta} = \beta \theta + (1-\beta)\bar{\theta}$
\EndFor
\end{algorithmic}
\end{algorithm}
\section{Experiments}
 This section thoroughly evaluates the Bottom Up Network (BUN) framework within two cooperative environments. Specifically, we consider the Cooperative Navigation task, introduced by~\cite{lowe2017multi}, and the Traffic Signal Control (TSC) task as described in~\cite{ault2021reinforcement}. To showcase the versatility of our approach, we extend our evaluation to encompass various adaptations of Cooperative Navigation and TSC. Detailed descriptions of these experimental environments can be found in the subsequent subsections. It is important to note that, for this study, we employ discrete actions across all environments. \par 
 \subsection{Benchmark Algorithms}
Our experiments are designed to compare the performance of BUN against several benchmark methods. Firstly, we evaluate BUN against independent Q-learning (Which we refer to as the Decentralized method), which serves as the baseline approach. In this baseline method, each agent operates independently without communication. To assess the efficacy of the sparse network architecture, we further compare BUN against RigL~\cite{evci2020rigging}. Additionally, we investigate the impact of performance and the utilization of Floating Point Operations per Second (FLOPS) compared to dense networks. This involves comparing BUN against two key configurations: (i) Centralized learning (i.e., Dense Model), which represents an ideal scenario where each agent has access to the entire global state, and (ii) DGN~\cite{jiang2019graph}, which utilizes an attention mechanism to enable communication between agents.
\subsection{Environments}
In the Cooperative Navigation environment, we deploy a scenario featuring N agents and N landmarks, where the overarching goal is for all agents to cover all the designated landmarks while avoiding collisions efficiently. To comprehensively evaluate the capabilities of BUN, we employ various variations of this environment. While we provide a brief overview, a more detailed description can be found in the supplementary material.\par
\textbf{Simple Spread (SS)}  In the Simple Spread task, N agents are tasked with reaching N landmarks. Each agent's observations encompass their position and the relative positions of the landmark assigned to them. The primary objective is for each agent to strategically position themselves to reach their landmarks. Notably, this environment serves as a baseline scenario where the performance of the Decentralized approach will be comparable to that of the centralized learning algorithm. This is because individual agents do not necessitate communication with their peers to achieve their objectives.\par
\textbf{SS with Communication (SS+C)}  In this task, we retain the same objective as Simple Spread, with 2 agents and 2 landmarks. However, each agent's observations now include their position and the relative positions of the other agent's designated landmark. Communication between agents becomes essential to successfully reach their respective landmarks as they need to share information about their landmarks. This configuration highlights the significance of agent-agent communication in achieving the mission. \par 
\textbf{SS with Cross Communication (SS+CC)}In this task, we introduce 3 agents and 3 landmarks. Each agent's observations still include their position and the relative positions of their assigned landmark. However, this environment has a unique twist: Agent 1 receives higher rewards when it occupies the landmark designated for Agent 3, and Agent 2 receives higher rewards when it occupies Agent 1 landmark. To optimize their returns, each agent must communicate strategically with the necessary agents to navigate toward the landmarks that yield higher rewards. This setup introduces an asymmetric communication requirement among the agents, differentiating it from the previous scenarios (2). This task would show the efficacy of our approach in establishing the necessary communication.  
\begin{figure*}
\vspace{-3mm}
     \centering
     \begin{subfigure}[b]{0.25\textwidth}
         \centering
         \includegraphics[width=\textwidth]{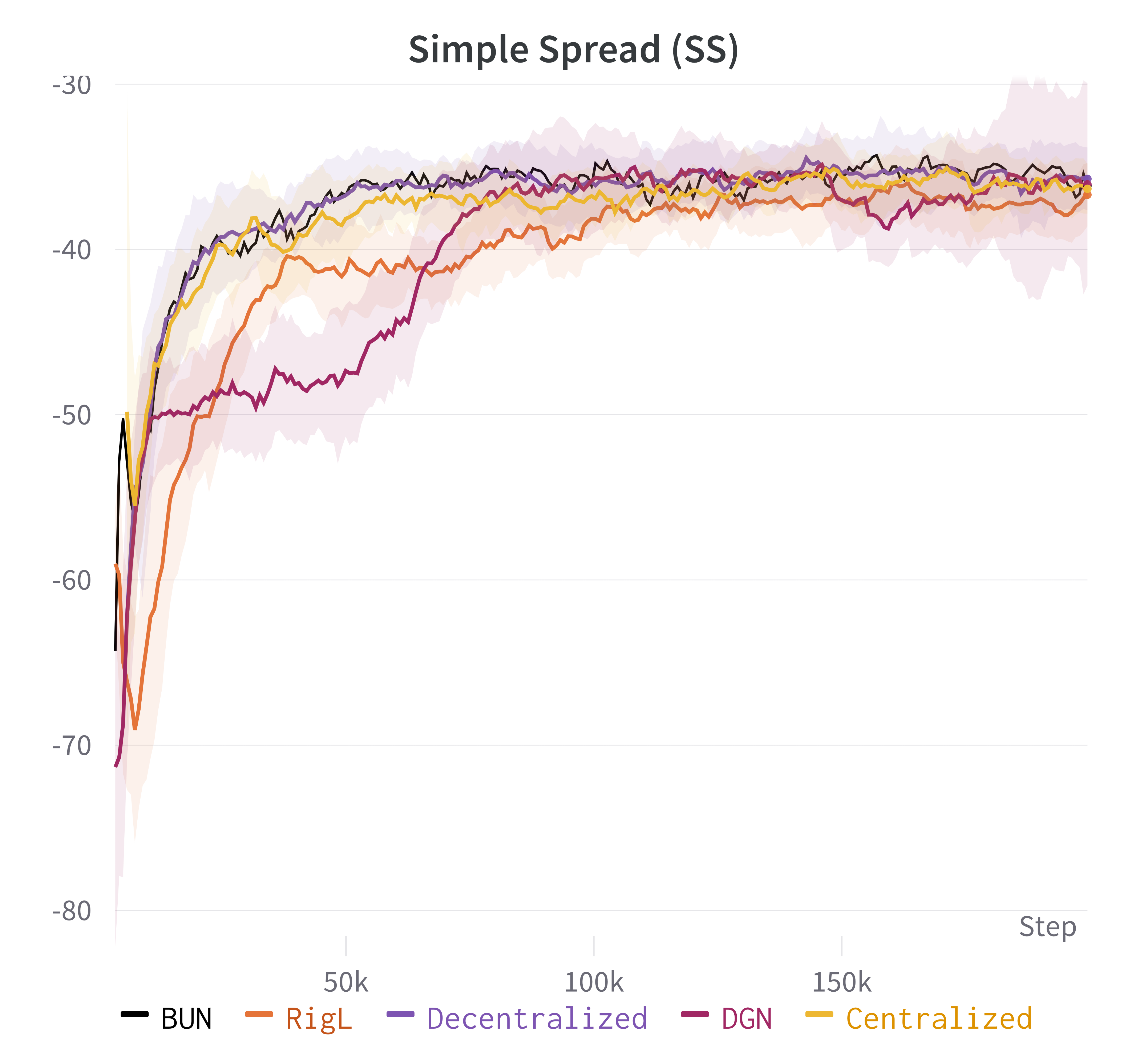}
         \label{SSplotsa}
     \end{subfigure}
     \hfill
     \begin{subfigure}[b]{0.25\textwidth}
         \centering
         \includegraphics[width=\textwidth]{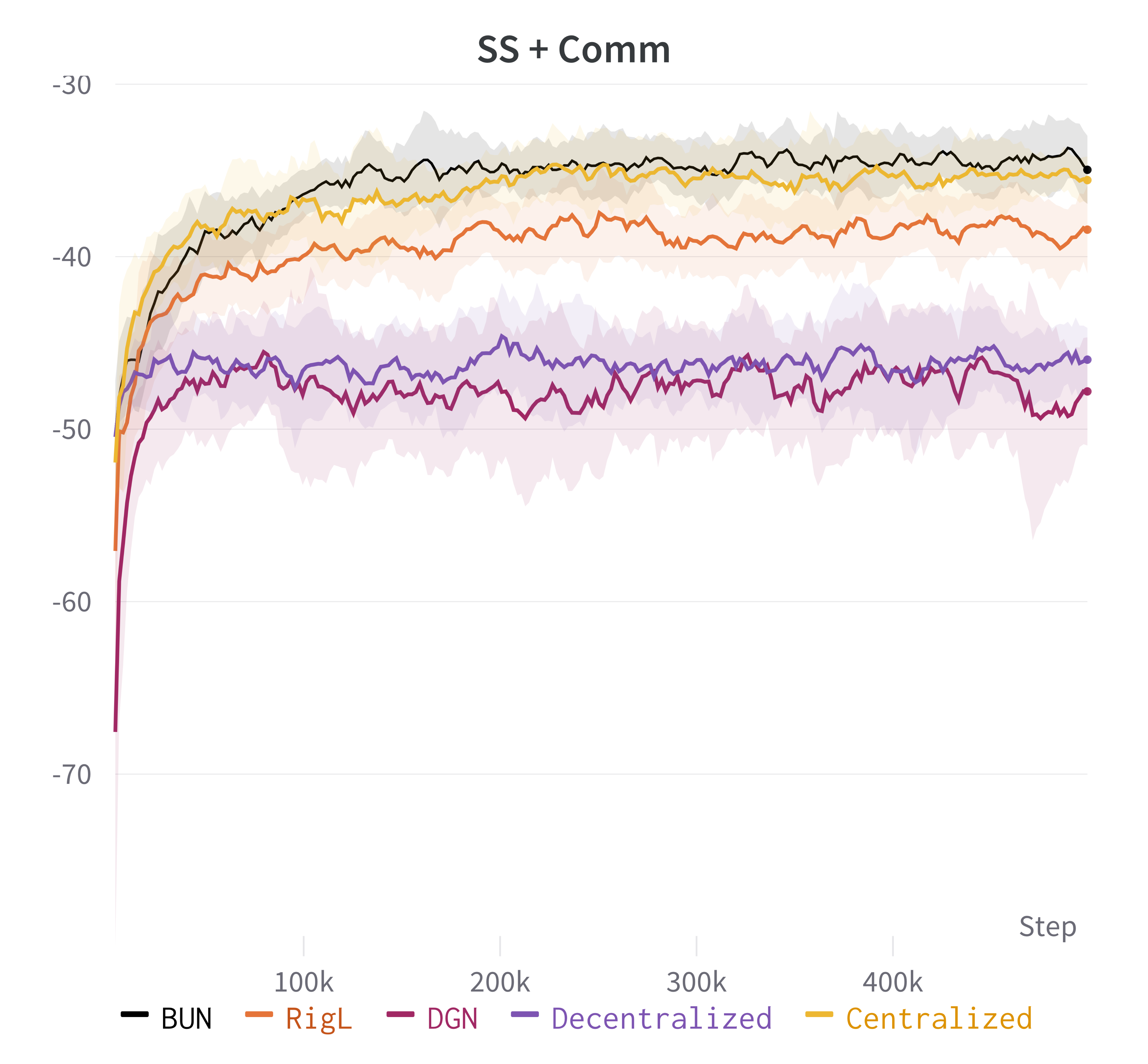}
         \label{SSplotsb}
     \end{subfigure}
     \hfill
     \begin{subfigure}[b]{0.25\textwidth}
         \centering
         \includegraphics[width=\textwidth]{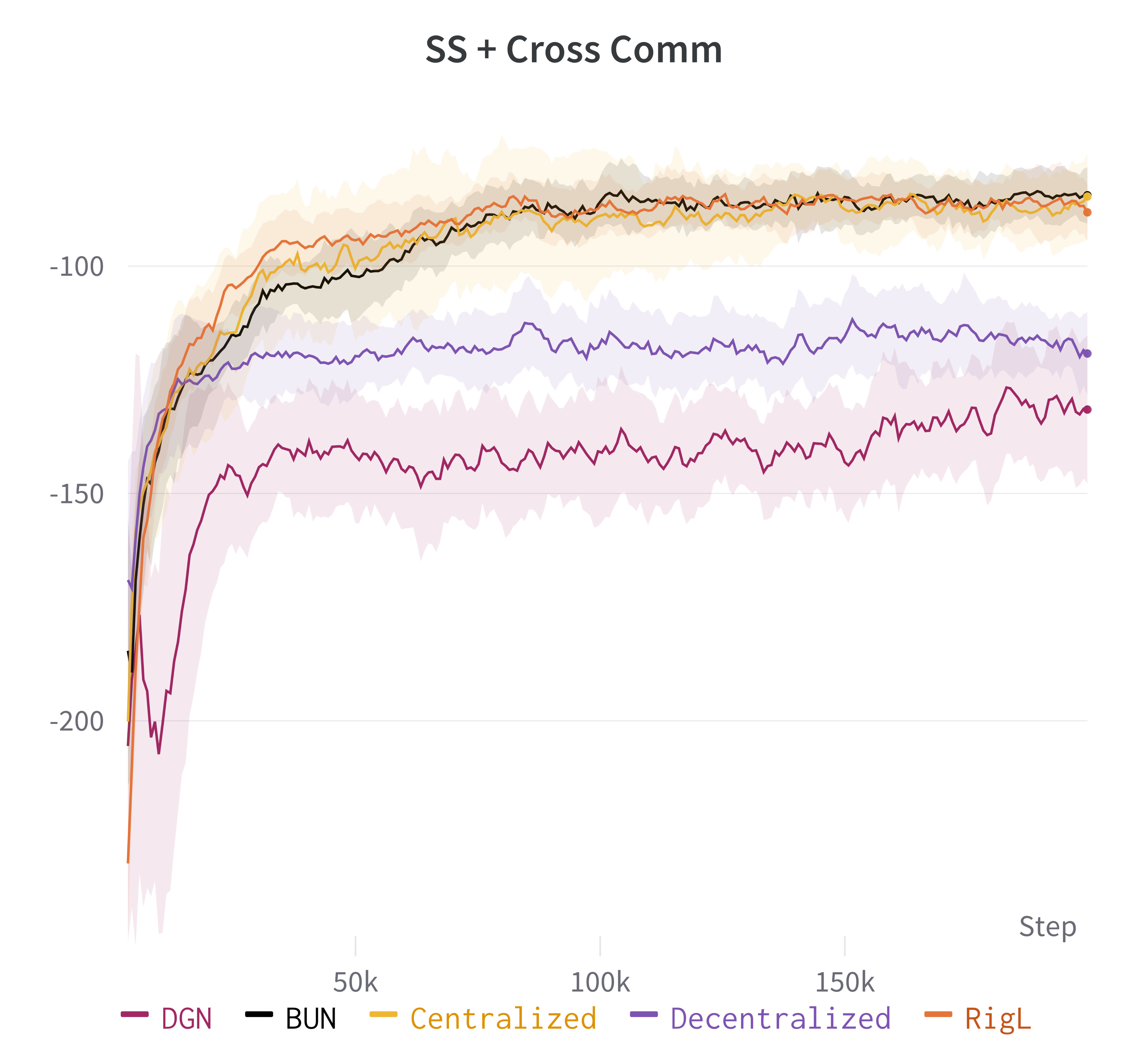}
         \label{SSplotsc}
     \end{subfigure}
        \caption{Learning curve during the training of Cooperative Navigation environments. Agents on  SS and SS + CC are trained for 20000 time-steps while on  SS+C  are trained for 500000 time-steps. The plots show the mean episode reward over 10 random seeds. }
        \label{SSplots}
\end{figure*}
\par\textbf{Traffic Signal Control}In the Traffic Signal Control environment, our objective is to assess the effectiveness of BUN in handling complex and dynamically changing traffic scenarios. In this simulated road network, each agent is a traffic signal controller at an intersection. An agent's observations comprise a one-hot representation of its current traffic signal phase (indicating directions for red and green lights) and the number of vehicles on each incoming lane at the intersection. At each time step, agents select a phase from a predefined set for the upcoming interval, typically set at 10 seconds. The overarching global objective is to minimize the average waiting time for all vehicles within the road network. To conduct these experiments, we use the framework~\cite{ault2021reinforcement} built on SUMO traffic simulator~\cite{lopez2018microscopic}. Specifically, we experimented with two different network configurations: a 2 x 2 grid network featuring four intersections and a smaller section of the Ingolstadt Region network comprising 7 intersections. Our traffic flow simulations encompassed a variety of scenarios, including both peak and off-peak periods, to emulate dynamic traffic patterns as given in~\cite{ault2021reinforcement}.
\subsection{Implementation Details}
In our training setup across all environments, our neural network architecture begins with three  (ReLU) layers, each sized at 18 times the number of agents. For the process of weight emergence, we gradually increase the network weights. This growth starts at step 10,000 $(T_{start})$ and continues until step 30,000 $(T_{end})$, with weight updates occurring at intervals of 1,000 steps $(\Delta T)$, and each update increases the number of weights by a factor of 3 ($k=3$). We introduce a deliberate delay to ensure that agents can learn from their observations before the weight growth begins. Weight growth commences only after 10,000 steps, and we have chosen a suitable $(\Delta T)$ to provide ample separation between weight updates, preventing the receipt of potentially misleading gradient signals. We also emphasize the importance of this step in our analysis, which is provided in the Supplementary Material. To maintain fairness in comparing different approaches, we ensure an equal number of neurons are employed across all methods and adjust hyperparameters accordingly. Detailed hyperparameter settings for each approach can be found in our Supplementary Material. \par 
In cooperative Navigation, we used 3 metrics to compare different methods: Mean Episode Reward (R), The average reward achieved by the team in an episode. Success Rate (S$\%$): In what percentage of episodes does the team achieve its objective? (Higher is better) Time (T): How many time steps does the team require to achieve its objective? (Lower is better). Each episode in the Cooperative Navigation environment lasts for a total of 25-time steps. Evaluation is carried out for each episode in the same seeded environment and illustrated in Figure~\ref{SSplots}. We then test the trained model on a newly set seeded environment for 25-time steps and show the results in Table~\ref{SSTable}. In the traffic environment, we used 2 metrics: Average Waiting time (Lower is better) and Average Trip time (Lower is better). Each episode is run for 3600 steps. We then test the trained model and provide performance results in Table~\ref{Traffictable}. In both environments, we use the metric floating-point operations (FLOPs) to show that our sparse method utilizes fewer arithmetic operations (Lower is better). 
\begin{figure*}
\vspace{-3mm}
     \centering
     \begin{subfigure}[b]{0.18\textwidth}
         \centering
         \includegraphics[width=\textwidth]{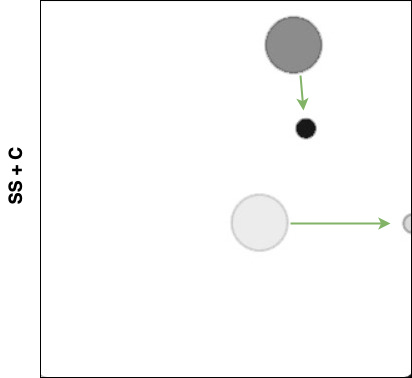}
         \label{SSplotsa}
     \end{subfigure}
     \hfill
     \begin{subfigure}[b]{0.16\textwidth}
         \centering
         \includegraphics[width=\textwidth]{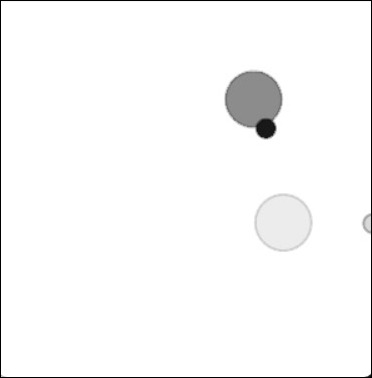}
         \label{SSplotsb}
     \end{subfigure}
     \begin{subfigure}[b]{0.16\textwidth}
         \centering
         \includegraphics[width=\textwidth]{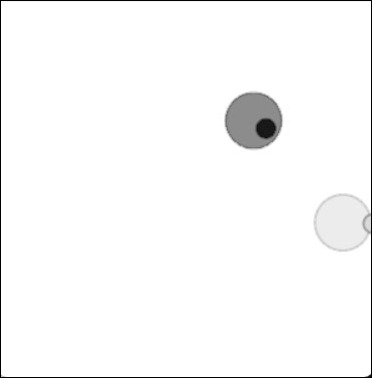}
         \label{SSplotsc}
     \end{subfigure}
     \hfill 
     \begin{subfigure}[b]{0.16\textwidth}
         \centering
         \includegraphics[width=\textwidth]{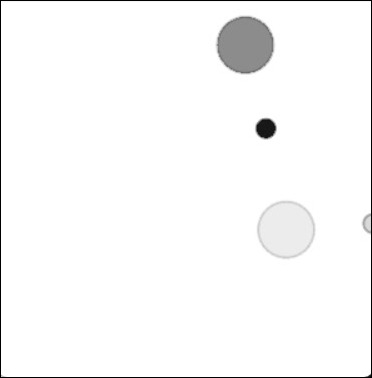}
         \label{SSplotsa}
     \end{subfigure}
     \begin{subfigure}[b]{0.16\textwidth}
         \centering
         \includegraphics[width=\textwidth]{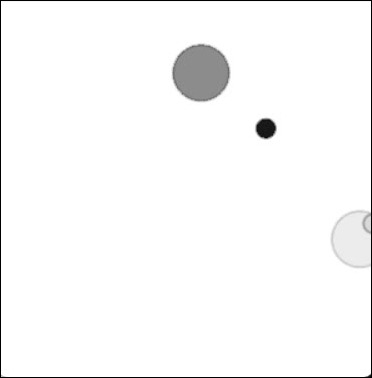}
         \label{SSplotsb}
     \end{subfigure}\\
      \begin{subfigure}[b]{0.18\textwidth}
         \centering
         \includegraphics[width=\textwidth]{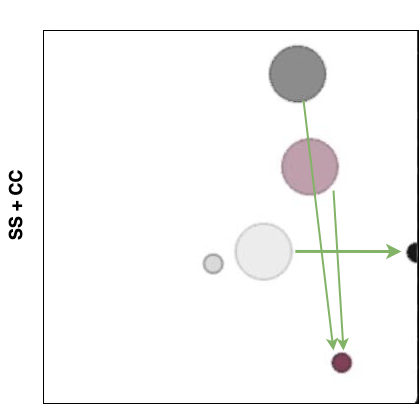}
         \caption{t = 0}
         \label{SSplotsa}
     \end{subfigure}
     \hfill
     \begin{subfigure}[b]{0.16\textwidth}
         \centering
         \includegraphics[width=\textwidth]{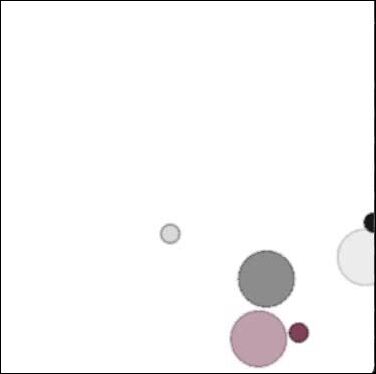}
         \caption{BUN (t = 6)}
         \label{SSplotsb}
     \end{subfigure}
     \begin{subfigure}[b]{0.16\textwidth}
         \centering
         \includegraphics[width=\textwidth]{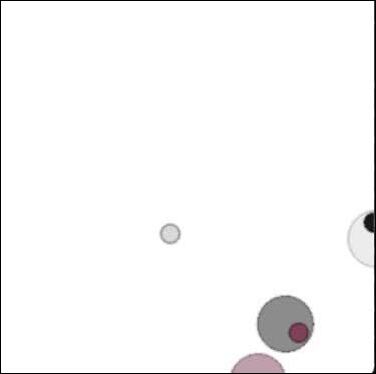}
         \caption{BUN (t = 10)}
         \label{SSplotsc}
     \end{subfigure}
     \hfill 
     \begin{subfigure}[b]{0.16\textwidth}
         \centering
         \includegraphics[width=\textwidth]{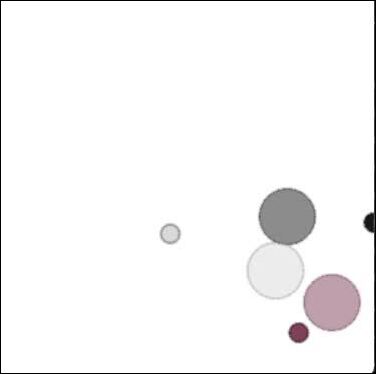}
         \caption{RigL (t = 10)}
         \label{SSplotsa}
     \end{subfigure}
     \begin{subfigure}[b]{0.16\textwidth}
         \centering
         \includegraphics[width=\textwidth]{Rigl10cc.jpg}
         \caption{RigL (t = 25)}
         \label{SSplotsb}
     \end{subfigure}
        \caption{Comparison between BUN (left) and RigL (right) on the Simple Spread with Communication (SS+C) and Simple Spread with Cross Communication (SS+C) environments at t = 0, 6, and 10 and t = 0, 10, and 25. Small circles indicate landmarks and Big circles indicate Agents. In SS+C, the white agent is assigned a white landmark, while the black agent is assigned a black landmark. In SS+CC, the white agent is penalized twice as a black agent to reach the black landmark, while red and black agents are assigned to the red landmark. The black agent is aggressive to reach the red landmark as it is penalized twice as the red agent to reach the red landmark. In both environments, the agents trained using BUN tried to grasp the information of their target landmarks from their fellow agents and reach the target landmarks.
On the other hand, the agents trained using RigL struggle to establish the connection between agents. In SS+C, the black agent reaches the black landmark, establishing that it only learned the local behaviour but did not establish the connection between the white agent. See the video for complete trajectories provided in the supplementary material.}
        \label{Simulationplots}
\end{figure*}
\begin{table*} 
\caption{In Cooperative Navigation environments, we assess the performance of various approaches in terms of Success Rate ($S\%$) and Time Steps (T) for all trained agents as they aim to reach their designated target landmark during testing. Additionally, we analyze the training cost, measured in FLOPs (Floating Point Operations), incurred by these different approaches during their training phase. It is to be noted that the BUN approach has a varying number of FLOPs as we grow the weights during the training phase. This table provides the average number of FLOPs utilized during training progress for the forward pass.}
\label{SSTable}
\begin{center}
\begin{small}
\begin{sc}
\begin{tabular}{l|ccc|ccc|ccc}
\toprule
&\multicolumn{9}{c}{Cooperative Navigation - Simple Spread (SS)}\\
&\multicolumn{3}{c}{SS}&\multicolumn{3}{c}{SS+C}&\multicolumn{3}{c}{SS+CC}\\
\midrule
Model& Flops & S$\%$ & T& Flops &  S$\%$  & T& Flops &  S$\%$  &T\\
\hline
Centralized & 4.2$e$3&100&10.25&4.2$e$3&100&10.36&9.5$e$3&100&16.12\\
 Decentralized & 2.7$e$3&100&9.75&2.7$e$3&0&25&3.2$e$3&0&25\\
 BUN &2.7$e$3&100&10.75&2.7$e$3&100&10.78&3.2$e$3&100&16.55\\
DGN &6.1$e$3&0&25&6.1$e$3&0&25&9.2$e$3&0&25\\
 RigL & 2.9$e$3&100&12.12&2.9$e$3&0&25&3.4$e$3&30&23.35\\
  \bottomrule
\end{tabular}
\end{sc}
\end{small}
\end{center}
\end{table*}
\begin{table*} \label{TrafficTable}
\caption{In Traffic Signal Control environments, we assess the performance of various approaches in terms of Average Waiting Time (Avg. Wait) and Average Trip Time (Avg. Trip Time) for all approaches. Additionally, we analyze the training cost, measured in FLOPs (Floating Point Operations), incurred by these different approaches during their training phase. To facilitate comparison, we normalize these training costs concerning the FLOPs utilized by the centralized approach, referred to as the Dense approach.  }
\label{Traffictable}
\begin{center}
\begin{small}
\begin{sc}
\begin{tabular}{l|c|c|c||c|c|c}
\toprule
&\multicolumn{3}{c||}{Grid 2$\times$2}&\multicolumn{3}{c}{Inglodast Corridor}\\
\midrule
Model& \multicolumn{1}{c}{FLOPS}&\multicolumn{1}{c}{Avg. Wait}&\multicolumn{1}{c||}{Avg. Trip Time}&\multicolumn{1}{c}{FLOPS}&\multicolumn{1}{c}{Avg. Wait}&\multicolumn{1}{c}{Avg. Trip Time}\\
\hline
Centralized & 2.6$e$6$(1x)$&2.41 & 69.61& 7.5$e$6$(1x)$&12.53&74.00\\
 Decentralized & 6.6$e$5$(0.25x)$&3.96 & 71.75&1$e$6$(0.14x)$&15.87&  77.36\\
 BUN &6.6$e$5$(0.25x)$ &2.25 & 69.35&1$e$6$(0.14x)$ &12.42& 73.86\\
 RigL &6.6$e$5$(0.25x)$ &2.39& 69.38& 1$e$6$(0.14x)$&13.83&76.27\\
  \bottomrule
\end{tabular}
\end{sc}
\end{small}
\end{center}
\vspace{-3mm}
\end{table*}
\subsection{Results}
\textbf{(SS)} In our experiments, we trained the models throughout 200,000-time steps. The learning curve for this training period in the Simple Spread environment is presented in Figure~\ref{SSplots}(a). In the centralized approach, where each agent has access to the observations of all other agents, optimal performance is achieved. However, given the problem's simplicity, where each agent can access its landmark information, demanding information from fellow agents becomes unnecessary. Consequently, the Decentralized approach converges to the same reward level as the Centralized approach. Similarly, the BUN, RigL, and DGN approaches converge to the same reward level as the Centralized approach. Nevertheless, there are differences in the utilization of computational resources, precisely the number of Floating Point Operations (FLOPs), among these approaches. As detailed in Table~\ref{SSTable}, DGN utilizes more FLOPs, followed by the Centralized approach, as both methods employ dense models. The increased FLOPs in DGN can be attributed to an attention mechanism in the approach. In contrast, Decentralized and BUN employ fewer FLOPs, while RigL utilizes slightly more FLOPs than BUN. However, despite these variations in FLOP utilization, all models perform similarly during testing, achieving comparable performance in both success rate ($S\%$)and Time (T), as demonstrated in Table~\ref{SSTable}. \par 
\textbf{(SS + C)} In this environment, we trained the model for about 500k time-steps until convergence. Figure~\ref{SSplots}(b) illustrates the learning for the training period. Since each agent has the observation of other agent's landmark position, there needs to be a necessary flow of information across agents to achieve the optimal reward. As expected, centralized and decentralized approaches achieve optimal and sub-optimal performances. BUN converges to Centralized performance, while RigL fails to reach the optimal reward. The same can be observed during evaluation as demonstrated in Table~\ref{SSTable}. The agents trained using centralized and BUN have the $100\%$ success rate in reaching their respective landmarks, while RigL and DGN have $0\%$ success rate. We illustrate the simulation in Figure~\ref{SSplots} to further demonstrate this. As shown in the figure, BUN agents learn the location of their landmarks and reach the landmarks, while only one agent learns the information about its landmark, and the other agent fails to learn the location of its landmark. We hypothesize that a primary reason for the failure of RigL in this setting is mainly due to the lack of a consistent weight emergent in the network of agents, as it iteratively tried to prune and grow the weights. BUN agents significantly gain from the weight initialization, where the agent's weight gets trained steadily, which aids in the consistent weight growth from a steady gradient signal across the agents. Surprisingly, DGN does not perform well. This is mainly due to the network size, as we chose to fix the smaller number of parameters in the network. Furthermore, as observed from the experiments in Simple Spread, DGN and Centralized approaches utilize more FLOPs, while BUN utilizes fewer FLOPs to achieve optimal performance. \par 
\begin{figure*}
\vspace{-5mm}
     \centering
     \begin{subfigure}[b]{0.26\textwidth}
         \centering
         \includegraphics[width=\textwidth]{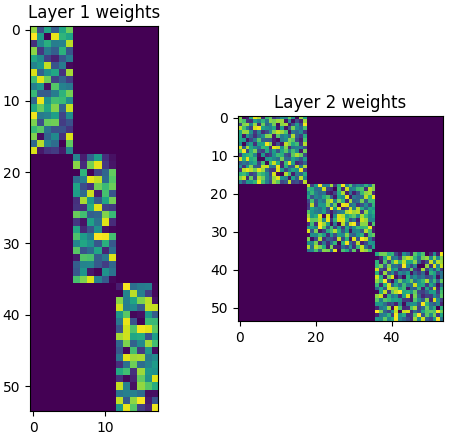}
         \caption{Initialization of Network}
         \label{SSplotsa}
     \end{subfigure}
     \hfill
     \begin{subfigure}[b]{0.26\textwidth}
         \centering
         \includegraphics[width=\textwidth]{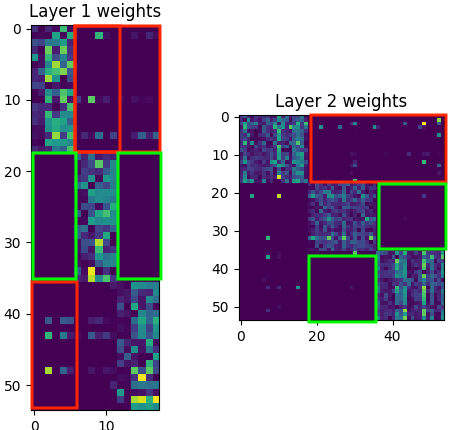}
         \caption{Trained weights using BUN}
         \label{SSplotsb}
     \end{subfigure}
     \hfill
     \begin{subfigure}[b]{0.26\textwidth}
         \centering
         \includegraphics[width=\textwidth]{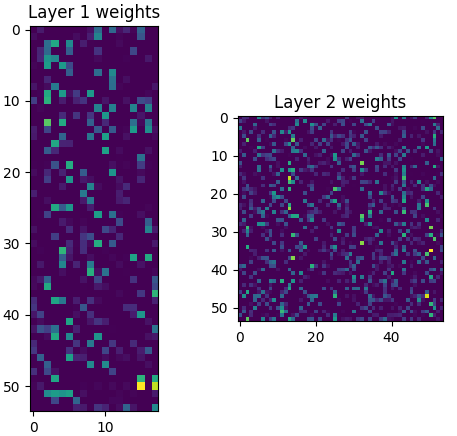}
         \caption{Trained weights using RigL}
         \label{Traffictable}
     \end{subfigure}
        \caption{In this comparison, we examine the training approaches of BUN and RigL within the context of the SS+CC environment. These figures showcase the evolution of neural network weights in both methods. In the BUN approach, training starts with local weight initialization (a), where agents operate independently. Agent observations follow a specific sequence, with black and white agents preceding red. The aim is to establish connections between agents (highlighted in red boxes) with no emergence of weights across agent red and agent white (green boxes). The weights in BUN emerge within a fixed budget (b = 30), as depicted in (b). Conversely, RigL exhibits a different pattern of weight emergence, as seen in (c). Unlike BUN, RigL introduces random weight connections. These structural weight emergence patterns shed light on the results presented in the accompanying table and the agent trajectories in Figure~\ref{SSTable}, highlighting each approach's distinct communication and coordination strategies.}
        \label{weightplots}
\end{figure*}
\textbf{(SS + CC)}
In this environment, we trained the models throughout 200,000 time steps, with each episode consisting of 25 time steps. Like the above two experiments, Centralized and Decentralized achieve optimal and suboptimal performance. Meanwhile, BUN and RigL converge to achieve the performance of a centralized approach. However, during the evaluation, RigL agents in some seeds failed to reach the assigned landmarks. We show an example in Figure~\ref{Simulationplots}, where agent black occupies the black landmark rather than the assigned red landmark and agent black blocks agent white from occupying its assigned landmark. This situation explains that the black agent is locally trained but has not developed the necessary communication links between agent black and Agent Red (since Agent Red has the locations of red landmarks). However, the agents trained using BUN achieve the tasks by reaching the assigned landmarks, and the red agent stays near the red landmark to minimize its distance penalty. Similar to the above experiments, BUN utilizes few FLOPs, and it is observed that, as the number of agents increases, the sparsity level is increased and would be essential in the case of large-scale systems. 
\par 
\textbf{Traffic Signal Control}
To validate our approach against the models mentioned earlier on Traffic Signal Control, we trained the models for 50,000-time steps and presented our results in Table~\ref{Traffictable}. In both scenarios (Grid 2$\times$2 and Inglodats 7), all approaches achieved similar results, while the BUN model exhibited marginal improvements in both metrics, although these improvements did not reach statistical significance. The complete training results are provided in the supplementary section.\par 
Our findings align with those published in~\cite{ault2021reinforcement}, highlighting that independent algorithms outperform coordinated control algorithms in realistic traffic scenarios. These experiments indicate that independent learning methods are sufficient for these applications and underscore the effectiveness of our approach in two key aspects: 1. We initiate the process with an independent setting, where we can achieve results similar to what independent learning algorithms typically accomplish. 2. Our approach provides the flexibility to remain in the independent setting. If necessary, we can facilitate the emergence of weights and the growth of connections across the junctions to enable essential communication. Detailed results in Table~\ref{Traffictable} demonstrate that in the Grid 2$\times$2 scenario, the Centralized approach consumes more FLOPs compared to the Decentralized, BUN, and RiGL models, which utilize only 25$\%$ of the FLOPs while achieving similar performance. Similarly, in the Inglodast Corridor, we achieved comparable performance to the Centralized approach using only 14$\%$ of the FLOPs.
As the number of junctions increases within a given scenario, the number of required FLOPs decreases significantly. This characteristic makes our approach particularly suitable for large-scale traffic network scenarios. For such scenarios, we can initially employ the independent setting. As needed, we can allow for the emergence of sparse connections across the junctions, providing the necessary communication and adaptability to traffic conditions.
\subsection{Robustness}
In this section, we aim to evaluate the robustness of sparse networks in the presence of noise. As previously explored in~\cite{graesser2022state} for a single agent setting, we assess the impact of progressively introducing noise into the observations and subsequently measuring its influence on a trained network for a Multi-Agent Setting. Specifically, we introduce Gaussian noise, sampled from a distribution with mean zero and variance $\sigma$, to each observation made by the agent. This experiment is conducted in two cooperative environments, SS+C and SS+CC, comparing the sparse network (BUN) and the centralized approach (i.e., the densest network). As depicted in Table~\ref{RobustTable}, it is evident that agents trained using the BUN model exhibit greater robustness to noise when compared to their centralized counterparts, a trend observed across both cooperative environments. In the SS+C environment, both models deliver strong performance under low noise conditions. However, as the noise level increases, centralized agents cannot cope with noise, often failing to reach their designated landmarks.
In contrast, the BUN agent maintains a higher level of robustness and performs relatively well even in the presence of elevated noise levels. Nevertheless, it is worth noting that even the BUN agents display a decrease in robustness at significantly high noise levels. This experiment suggests sparse networks, such as BUN, exhibit greater resilience to observational noise than dense networks, such as the centralized approach.
\begin{table} 
\caption{To assess the robustness of networks trained using both the BUN and Centralized (Dense) approaches, we conducted tests involving introducing Gaussian noise to the observations. We sampled noise from a Gaussian distribution $\sim \mathcal{N}(0, \sigma) $, where $\sigma \in [0, 0.5]$, denoted as $\sigma$. Notably, we present the results at four specific data points}
\label{RobustTable}
\begin{center}
\begin{small}
\begin{sc}
\begin{tabular}{l|c|c|c}
\toprule
&variance&\multicolumn{2}{c}{Sucess Rate ($S\%$)}\\
Environment&($\sigma$)& Centralized&BUN\\
\hline
&0&100&100\\
SS&0.1&100&100\\
(C)&0.3&0&100\\
&0.5&0&0\\
\hline 
&0&100&100\\
SS&0.1&0&100\\
(CC)&0.3&0&75\\
&0.5&0&0\\
  \bottomrule
\end{tabular}
\end{sc}
\end{small}
\end{center}
\vspace{-3mm}
\end{table}

\section{Conclusion}
In this study, we introduced BUN, an efficient algorithm for training sparse neural networks in multi-agent environments. Notably, BUN outperforms existing dense and sparse training algorithms within prescribed computational budgets. Our approach demonstrates its value across three critical scenarios: first, by enhancing the performance of multi-agents through improved communication; second, by optimizing the utilization of available computational resources; and third, by serving as an initial step in interpreting the underlying topology among agents. In essence, BUN offers a versatile solution that elevates accuracy while respecting resource constraints, fosters coordination among agents, and unveils essential insights into the structure of multi-agent systems, thus making a substantial contribution to the field.







\bibliographystyle{ACM-Reference-Format} 
\bibliography{sample}

\appendix
\newpage 
\section{Supplementary Material}
\subsection{Analysis on Weight Emergence with $\Delta T$}
To gain a deeper mathematical understanding, we examine the optimization process. Consider the objective function value \(\mathcal{L}^{k+1}\) after the \(k\)-th iteration. We update \(\theta_{ij}^{k}\) in the direction of the negative gradient of \(J\) concerning \(\theta_{ij}\) while keeping all other weights fixed. Let \(\nabla_{ij} J\) denote the gradient of \(J\) with respect to \(\theta_{ij}\). 
Since \(\mathcal{L}\) is a continuously differentiable loss function, we can use the first-order Taylor expansion around \(\theta^{k}\) to approximate \(\mathcal{L}^{k+1}\) as follows:
\begin{equation} \label{taylorexpression}
    \mathcal{L}^{k+1} \approx \mathcal{L}^{k} + \nabla \mathcal{L}^{k} \cdot (\theta^{k+1} - \theta^{k})
\end{equation}
Consider the component corresponding to $\nabla \mathcal{L}^{k} \cdot (\theta^{k+1} - \theta^{k})$ in \eqref{taylorexpression} and expand it as follows:
\begin{align*}
    \nabla \mathcal{L}^{k} \cdot (\theta^{k+1} - \theta^{k}) \leq & \sum_{(i,j) \in \mathcal{S}^k}\nabla_{ij} \mathcal{L}^{k}\left( \theta_{ij}^{k+1} - \theta_{ij}^{k} \right) \\ 
    &+  \sum_{(i,j) \in \mathcal{S}_c^k}\nabla_{ij} \mathcal{L}^{k}\left( \theta_{ij}^{k+1} - \theta_{ij}^{k} \right) 
\end{align*}
Now, substitute the above expression back in~\eqref{taylorexpression} 
\begin{align} \label{changeinnorm}
    \mathcal{L}^{k+1} - \mathcal{L}^{k}  \leq &\sum_{(i,j) \in \mathcal{S}^k}\nabla_{ij} \mathcal{L}^{k}\left( \theta_{ij}^{k+1} - \theta_{ij}^{k} \right) \nonumber \\ 
    &+  \sum_{(i,j) \in \mathcal{S}_c^k}\nabla_{ij} \mathcal{L}^{k}\left( \theta_{ij}^{k+1} - \theta_{ij}^{k} \right) 
\end{align}
where $\mathcal{S}$ is a space of indices of the non-sparse weights, while $\mathcal{S}_c$ is a space of the sparse weights. In this form, the expression emphasizes that the change in the loss depends on the contributions from both the non-sparse weights $\{\theta_{ij}|(i, j) \in S^k\}$ and the sparse weights $\{\theta_{ij}|(i, j) \in S^k_c\}$, and each contribution is proportional to the corresponding gradient multiplied by the change in the element value. However, if the weights are trained properly, we can use the fact that $\nabla_{ij} \mathcal{L}^{k} = 0\; \forall\; (i, j) \in \mathcal{S}^k$. Then the expression~\eqref{taylorexpression} reduces to
\begin{align}
    \mathcal{L}^{k+1} - \mathcal{L}^{k} & \leq \sum_{(i,j) \in \mathcal{S}_c^k}\nabla_{ij} \mathcal{L}^{k}\left( \theta_{ij}^{k+1} - \theta_{ij}^{k} \right)
\end{align}
  In this context, while the change in the loss depends on the gradients of sparse weights, the pivotal influence of tuning becomes apparent. By training the non-sparse weights, our algorithm prioritizes sparse weights efficiently and optimizes the values of non-sparse weights. For this reason, we ensure a gap in each weight growth update. This strategy ensures that the change in the loss aligns closely with the optimization objective, making our approach highly effective in achieving superior performance while having a good gradient signal. 
\subsection{Hyperparametrs}
For the environments, Cooperative Navigation and Traffic Signal Control, we use a set of hyperparameters to achieve the results shown in our paper. We provide the hyperparameters used for Cooperative Navigation in Table~\ref{hyperparametersmpe} and Traffic Signal Control in Table~\ref{hyperparameterstraffic}. It is to be noted that the hyperparameters for Centralized and Decentralized approaches are the same as BUN.
\begin{table*} 
\caption{Hyperparameters of BUN, RigL and DGN. Where $N$ is the number of agents.}
\label{hyperparametersmpe}
\begin{center}
\begin{small}
\begin{sc}
\begin{tabular}{c|c|c|c}
\toprule
Hyperparameter & BUN & RigL & DGN\\
\midrule
discount ($\gamma$)&1024&1024&1024 \\ 
Buffer Capacity &1$e$6&1$e$6&1$e$6 \\ 
$\beta$&&& \\ 
decay($\epsilon$)&0.1&0.1&0.1 \\ 
optimizer &Adam&Adam&Adam \\ 
learning rate &1$e$-4&1$e$-4& 1$e$-4\\ 
layer type &MLP&MLP&MLP \\ 
$\#$ of layer &3&3&3 \\ 
$\#$ of units &(18*N, 18*N, 18*N)&(18*N, 18*N, 18*N)&(18, 18, 18) \\  
Activation type &ReLU&ReLU&ReLU \\ 
Weight Initialization &BUN&random normal&random normal \\ 
$\#$ of neighbors &-&-&all \\ 
$T_{start}$&10k&5k&- \\ 
$T_{end}$&30k&0.75*time&- \\ 
$\Delta T$&1000&100& -\\ 
\hline
\end{tabular}
\end{sc}
\end{small}
\end{center}
\end{table*}
\begin{table*} 
\caption{Hyperparameters of BUN, RigL and DGN. Where $N$ is the number of agents.}
\label{hyperparameterstraffic}
\begin{center}
\begin{small}
\begin{sc}
\begin{tabular}{c|c|c|c}
\toprule
Hyperparameter & BUN & RigL & DGN\\
\midrule
Batch Size ($\gamma$)&64&64&64 \\ 
Buffer Capacity &1$e$6&1$e$6&1$e$6 \\ 
$\beta$&&& \\ 
decay($\epsilon$)&0.1&0.1&0.1 \\ 
optimizer &Adam&Adam&Adam \\ 
learning rate &1$e$-4&1$e$-4& 1$e$-4\\ 
layer type &MLP&MLP&MLP \\ 
$\#$ of layer &3&3&3 \\ 
$\#$ of units &(256*N, 256*N, 256*N)&(256*N, 256*N, 256*N)&(256, 256, 256) \\  
Activation type &ReLU&ReLU&ReLU \\ 
Weight Initialization &BUN&random normal&random normal \\ 
$\#$ of neighbors &-&-&all \\ 
$T_{start}$&10k&5k&- \\ 
$T_{end}$&30k&0.75*time&- \\ 
$\Delta T$&1000&100& -\\ 
\hline
\end{tabular}
\end{sc}
\end{small}
\end{center}
\end{table*}
\begin{figure*}
     \centering
     \begin{subfigure}[b]{0.48\textwidth}
         \centering
         \includegraphics[width=\textwidth]{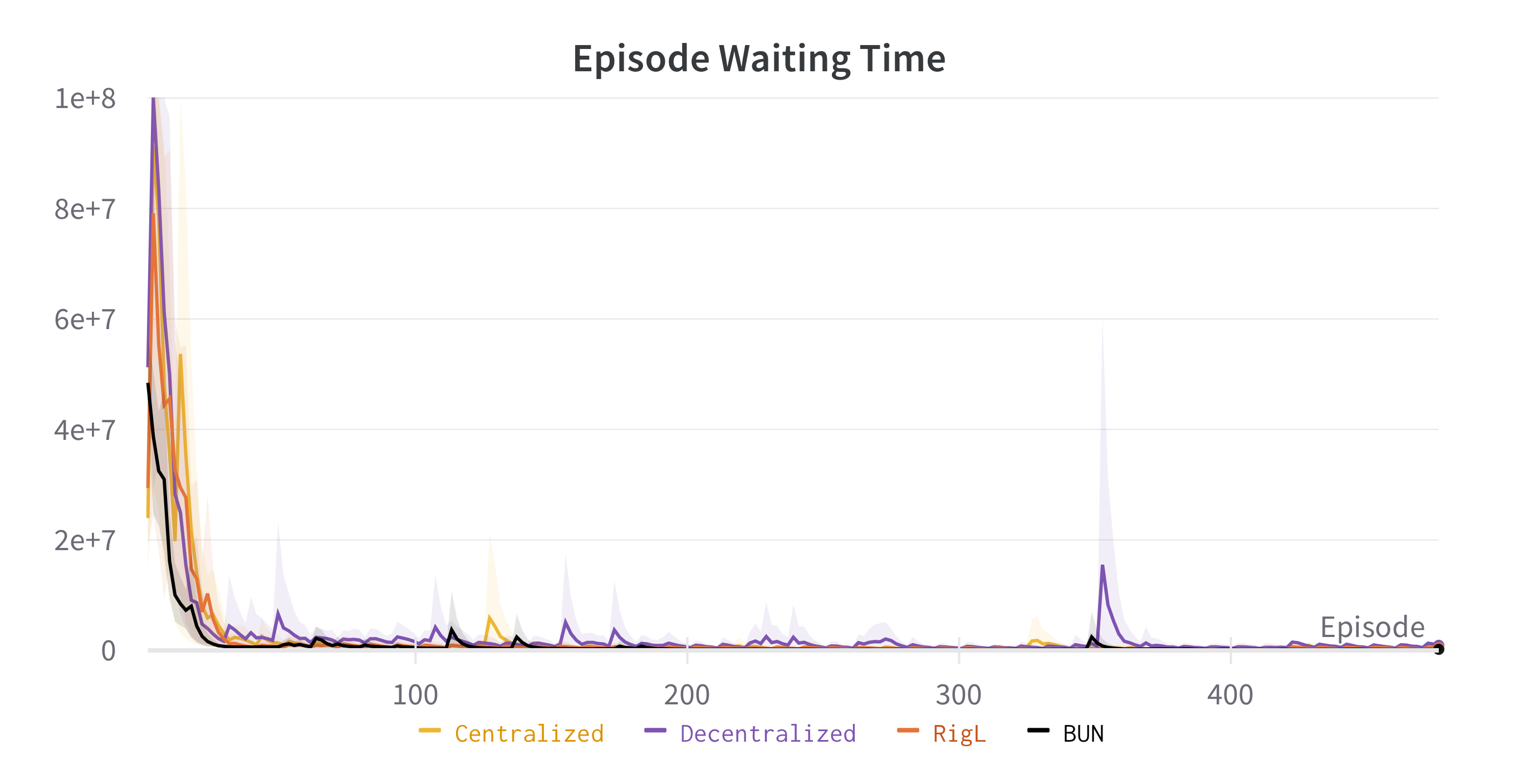}
         \label{SSplotsa}
     \end{subfigure}
     \hfill
     \begin{subfigure}[b]{0.48\textwidth}
         \centering
         \includegraphics[width=\textwidth]{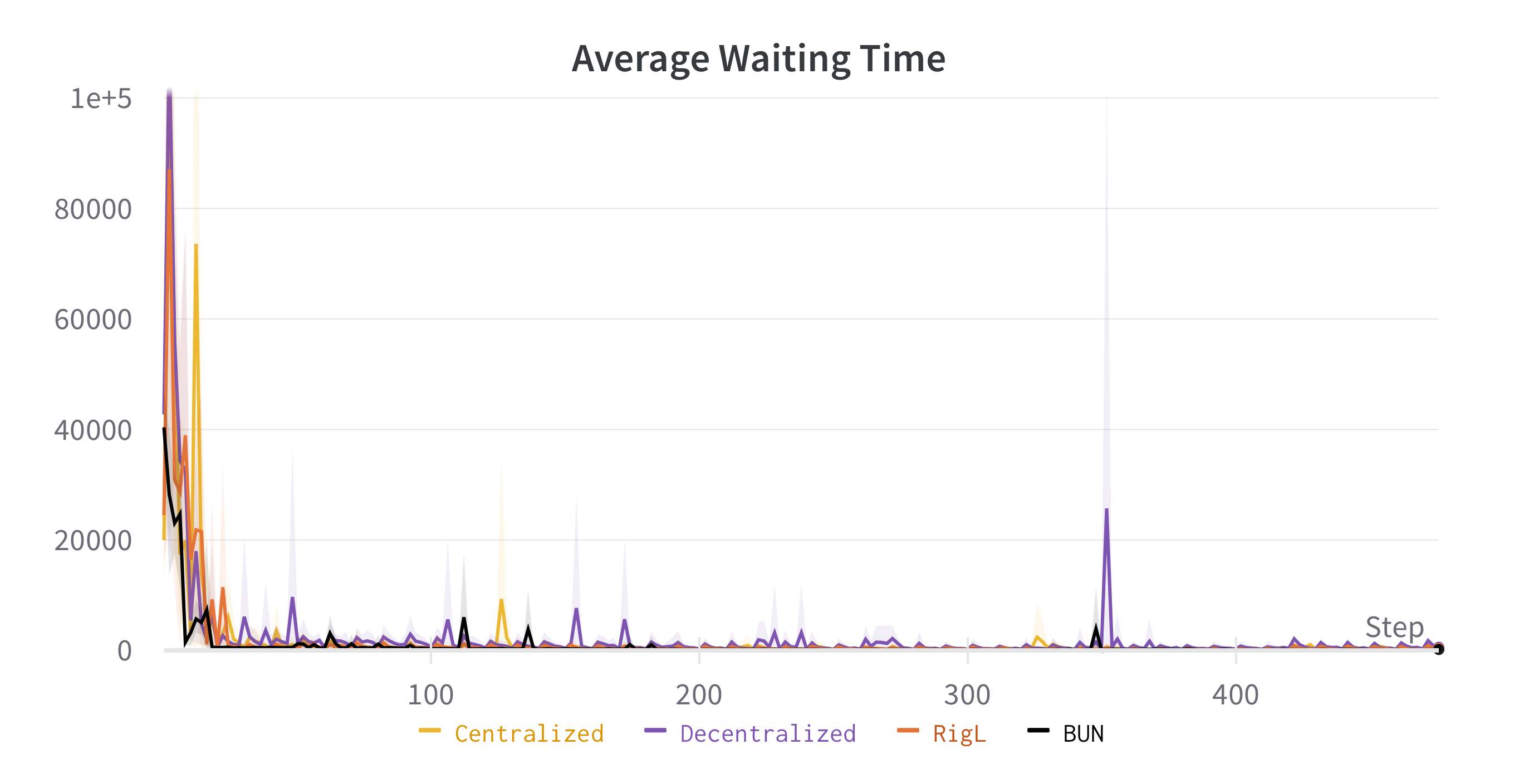}
         \label{SSplotsb}
     \end{subfigure}
        \caption{Learning curve during the training of Grid 2$\times$ environment. The plots show the Episode waiting time and Average Waiting Time of Vehicle in the grid network.}
        \label{trafficplotsconvergenceg4}
\end{figure*}
\begin{figure*}
     \centering
     \begin{subfigure}[b]{0.48\textwidth}
         \centering
         \includegraphics[width=\textwidth]{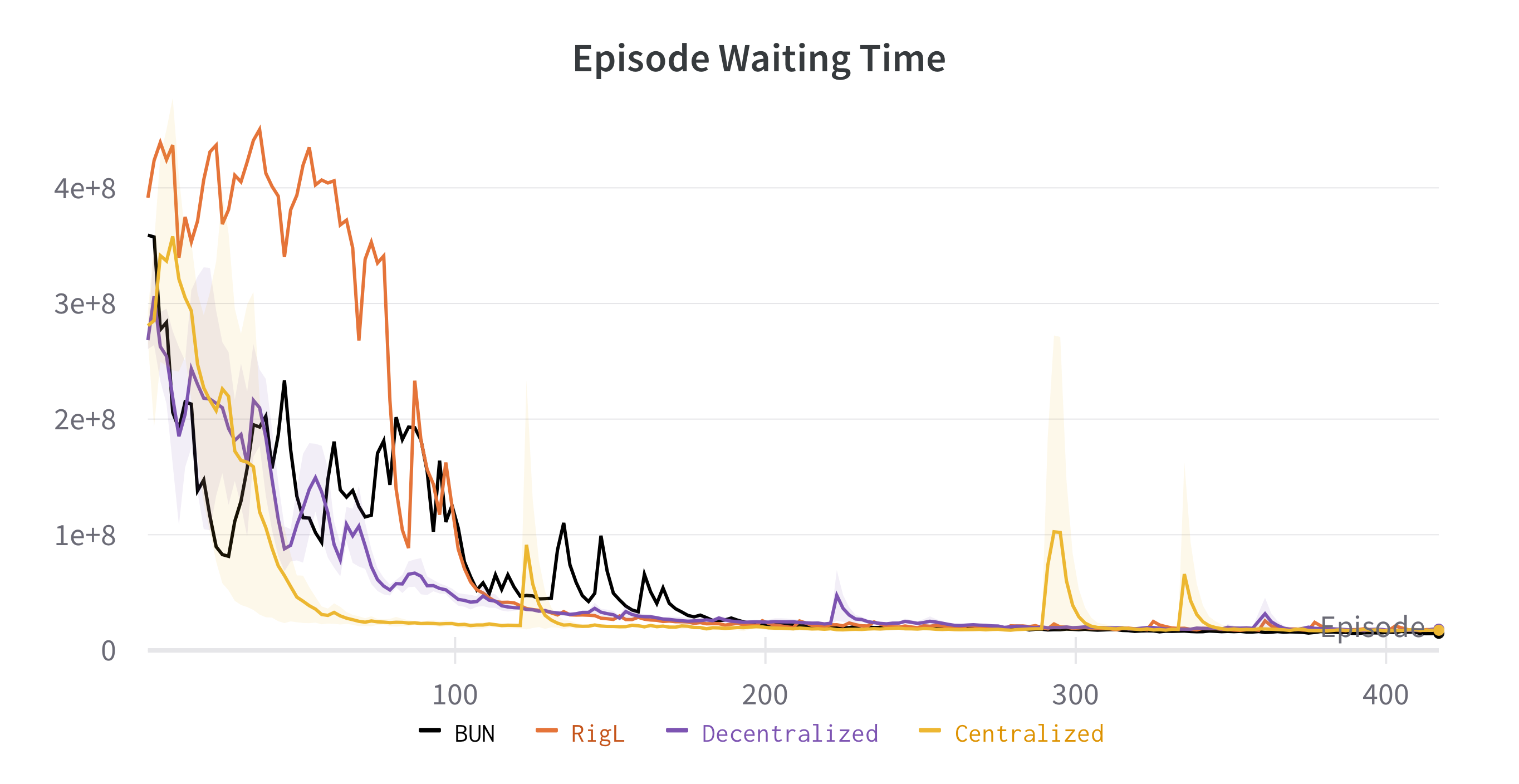}
         \label{SSplotsa}
     \end{subfigure}
     \hfill
     \begin{subfigure}[b]{0.48\textwidth}
         \centering
         \includegraphics[width=\textwidth]{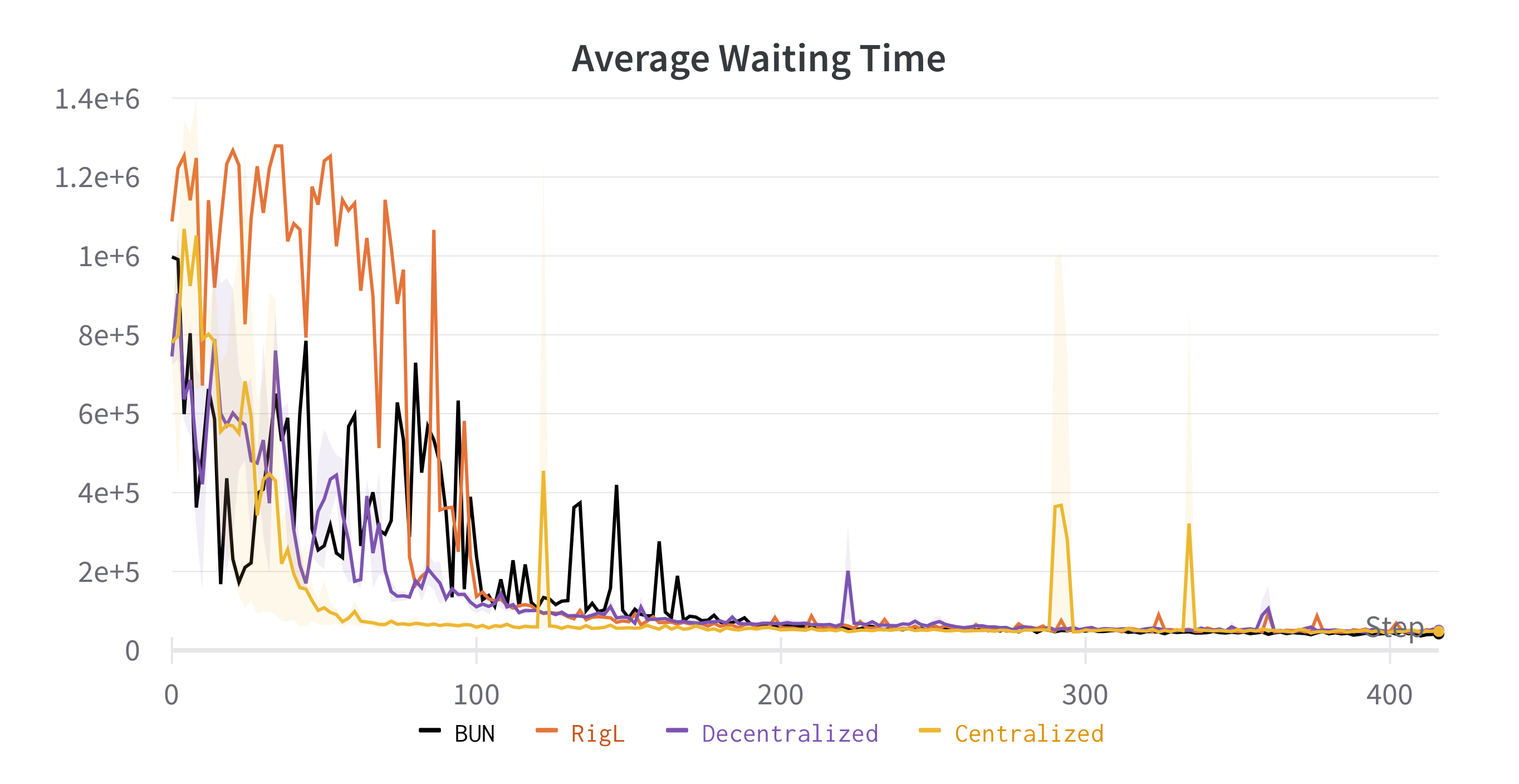}
         \label{SSplotsb}
     \end{subfigure}
        \caption{Learning curve during the training of Inglodast Corridor environment. The plots show the Episode waiting time and Average Waiting Time of Vehicle in the grid network.}
        \label{trafficplotsconvergencei7}
\end{figure*}
\begin{figure*}
     \centering
     \begin{subfigure}[b]{0.45\textwidth}
         \centering
         \includegraphics[width=\textwidth]{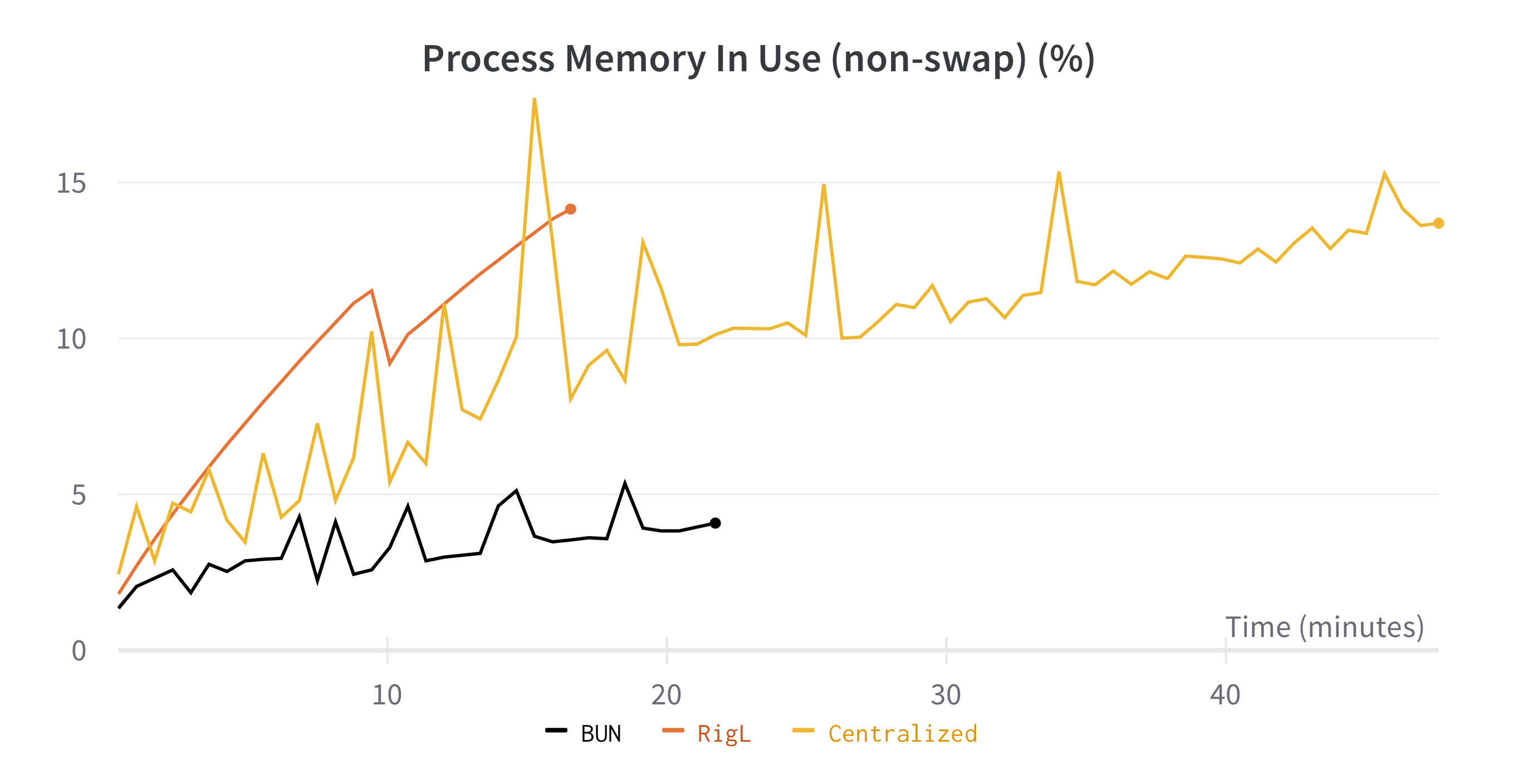}
         \label{SSmema}
     \end{subfigure}
     \hfill
     \begin{subfigure}[b]{0.45\textwidth}
         \centering
         \includegraphics[width=\textwidth]{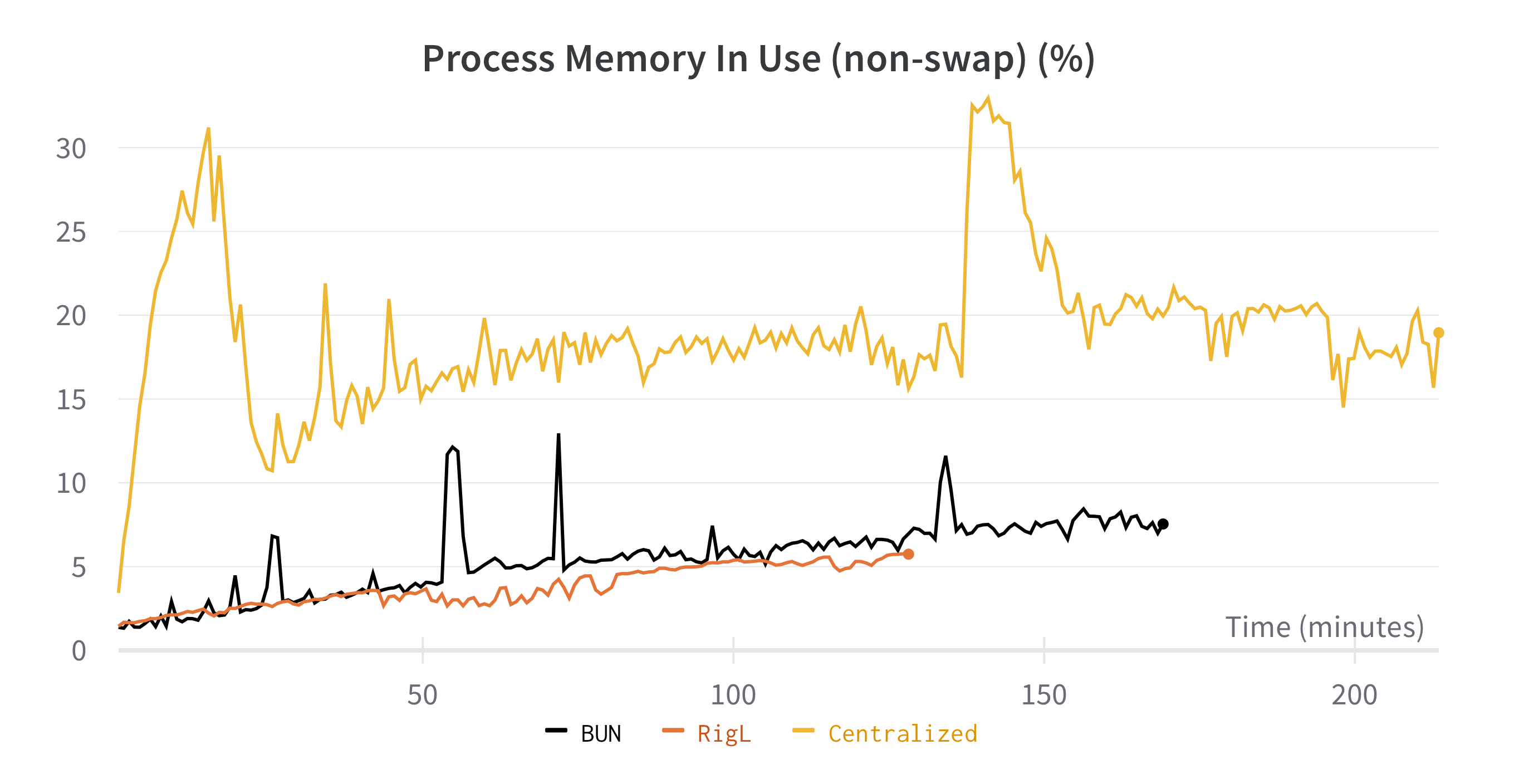}
         \label{SSmemb}
     \end{subfigure}
     \begin{subfigure}[b]{0.45\textwidth}
         \centering
         \includegraphics[width=\textwidth]{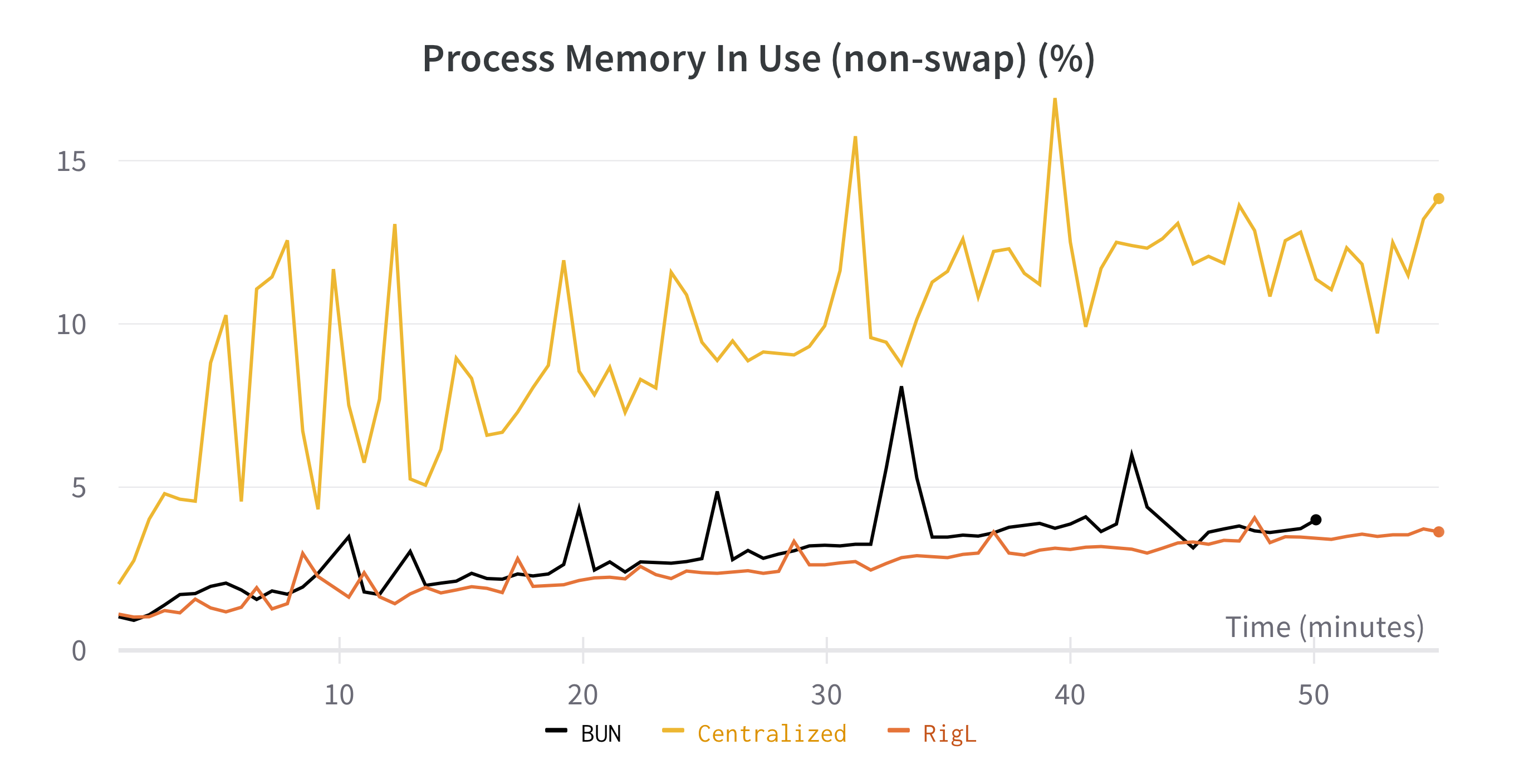}
         \label{SSmemc}
     \end{subfigure}
        \caption{Learning curve during the training of Inglodast Corridor environment. The plots show the Episode waiting time and Average Waiting Time of Vehicle in the grid network.}
        \label{memorycurves}
\end{figure*}
\subsection{Experimental Settings}
All benchmarks are performed on Apple M1, $2020$ (maci64) with 16 GB RAM. In the context of cooperative navigation, agents operating in an environment receive rewards based on their actions. These rewards serve two primary purposes: penalizing agent collisions with a reward of -1 and encouraging agents to minimize the distance between themselves and their assigned landmarks. This reward structure promotes collision avoidance and efficient goal achievement. Notably, the reward distribution is agent-specific, meaning that different agents receive rewards tailored to their unique objectives. In SS+CC, Agent 1 is incentivized to occupy Landmark 3 with a reward twice the negative distance between its current location and Landmark 3. Agent 1 also incurs a -1 penalty for collisions, ensuring it actively avoids collisions with other agents. Similarly, Agent 2 is encouraged to occupy Landmark 1 through a similar reward structure, receiving a -1 penalty for collisions. This approach ensures that each agent has a clear goal and incentive to achieve it while taking active measures to avoid collisions with other agents. In a traffic signal control environment, the reward function differs. Here, the rewards are associated with signal control at traffic junctions. The measure of success is the average queue length at these junctions. The goal is to minimize queue lengths, which, in turn, reduces the average waiting time for vehicles at these intersections. This approach aligns with the objectives of traffic management using reinforcement learning, where agents (in this case, traffic signals) are trained to actively optimize traffic flow while avoiding situations that lead to excessive vehicle queues and congestion. 
\subsection{Additional Results}
In this section, we provide the training curves for traffic signal control. In this section, we present the training curves for traffic signal control. Figure~\ref{trafficplotsconvergenceg4} demonstrates that in the 2$\times$2 grid scenario, all the approaches converge at a similar rate. However, the centralized approach converges faster in the Ingolstadt corridor region, as depicted in Figure~\ref{trafficplotsconvergencei7}. Compared to the centralized approach, both BUN and RigL and the decentralized approach exhibit slower convergence due to their reliance on learning traffic patterns from local observations. Notably, when considering partial observations, BUN outperforms RigL and the Decentralized approach regarding convergence speed.
Additionally, we provide insights into the various approaches memory utilization performance, as shown in Figure~\ref{memorycurves} and Sparsity level of networks in Table~\ref{sparsity}. The Figure~\ref{memorycurves} highlights that BUN and RigL utilize a lower percentage of memory than the centralized approach. This observation serves as supporting evidence for the efficiency of using fewer FLOPs (floating-point operations) during training.
\begin{table*} 
\caption{The percentage of sparsity varies across different approaches in Cooperative Navigation and Traffic Signal Control environments, with (N) representing the number of agents in the environment. It is evident that as the number of agents increases, the percentage of sparsity also increases. This trend highlights the scalability of sparse approaches in handling larger numbers of agents. Notably, the sparsity level of the BUN approach initially matches that of the decentralized approach but gradually increases per the allocated budget, eventually reaching a level comparable to that of RigL}
\label{sparsity}
\begin{center}
\begin{small}
\begin{sc}
\begin{tabular}{l|ccccc}
\toprule
&\multicolumn{5}{c}{Sparisty ($\%$)}\\
&SS(N=2)&SS+C(N=2)&SS+CC(N=3)&Grid 2$\times$2(N=4) &Ing. corridor(N=7)\\
\midrule
Centralized &0&0&0&0&0\\
 Decentralized &50&50&66.66&75&85.71\\
 BUN &57&57&64.4&75&85.71\\
DGN &0&0&0&0&0\\
 RigL &57&57&64.4&75&85.71\\
  \bottomrule
\end{tabular}
\end{sc}
\end{small}
\end{center}
\end{table*}

\end{document}